\RequirePackage{fix-cm}
\documentclass[pdftex,twocolumn,epjc3]{svjour3}       
\smartqed  
\usepackage{graphicx}
\begin{document}

\newcommand{\prd} {Phys. Rev. D}
\newcommand{\apj} {ApJ}
\newcommand{\mnras} {Mon. Not. Roy. Ast. Soc.}
\newcommand{\apss} {Astrophys. and Space Sci.}

\title{Frequentist Model Comparison Tests of  Sinusoidal Variations in Measurements of  Newton's Gravitational Constant
}


\author{Shantanu Desai \thanksref{addr1,e1}}

\thankstext{e1}{shntn05@gmail.com}           
\institute{Department of Physics, IIT Hyderabad, Kandi, Telangana-502285, India \label{addr1} 
}

\date{Received: date / Accepted: date}

\maketitle

\begin{abstract}
In 2015, Anderson et al~\cite{Anderson15a} have claimed to find evidence  for periodic sinusoidal variations (period=5.9 years)  in measurements of Newton's gravitational constant. These claims have been disputed by Pitkin~\cite{Pitkin15}.  Using Bayesian model comparison, he argues  that a model with an unknown Gaussian noise component is favored over any periodic variations by more than $e^{30}$. We re-examine the claims of Anderson et al~\cite{Anderson15a}  using
frequentist model comparison tests, both with and without errors in the measurement times. Our findings lend support to Pitkin's claim  that a constant term along with an unknown systematic offset provides a better fit to the measurements of Newton's constant, compared to any sinusoidal variations.

\end{abstract}

\section{Introduction}
\label{intro}
In 2015, Anderson et al~\cite{Anderson15a} have found evidence   for periodicities in the measured values of  Newton's gravitational constant ($G$) (using data compiled  in ~\cite{Schlamminger}) with a period of 5.9 years. They also noted that similar variations have been seen in the length of a day~\cite{Holme}, and hence there is a possible correlation between the two. However, these results have been disputed by Pitkin~\cite{Pitkin15} (hereafter P15). In P15, he examined four different models for the observed values of $G$ and  showed using Bayesian model comparison  tests that the logarithm of the Bayes factor for a  model with constant offset and Gaussian noise compared to sinusoidal  variations in $G$ is  about 30 (See Table 1 of P15). The analysis was done by considering both a uniform prior and  a Jeffreys prior for the parameters. Therefore, from his analysis the data is better fit  by a constant offset and an unknown Gaussian noise component. Thereafter, Anderson et al responded to this in a short note~\cite{Anderson15b}, pointing out that they were unable to replicate the claims in P15 using minimization of L1 norm,  and they stand by their original claims  of sinusoidal variations in $G$. Another difference between the analysis done by P15 and by Anderson et al is that P15 marginalized over the errors in measurement times of $G$,  whereas these errors in measurement times  ignored by Anderson et al. A periodicity search was also done by Schlamminger et al~\cite{Schlamminger}, which involved minimization of both the L1 and L2 norms. They also argue that a sinusoidal variation with a period of 5.9 years provides a better fit than a straight line. However, the chi-square probabilities  for all their models are very small (See Table III of ~\cite{Schlamminger}).

To resolve this imbroglio, we re-analyze the same data using Maximum Likelihood analysis (both with and without errors in 
the measured values of $G$) and do frequentist model comparison tests between different models.   Therefore, our analysis is complementary to that of Anderson~\cite{Anderson15a} and Schlamminger~\cite{Schlamminger}.
\section{Analysis summary}
\label{sec:1}
The sinusoidal model to which we fit the measurements of $G$ (obtained using a set of measurements $y_i$)
at times $t_i$ is given by~\cite{Anderson15a}:
\begin{equation}
y_i= A\sin[\phi_0 + 2\pi (t_i/P)] + \mu_G,
\label{eq}
\end{equation}
where $\mu_G$ is a constant offset, $P$ is the period of the sinusoid, $A$ is the amplitude of the modulation,  and $\phi_0$ its phase.

Similar to P15, we examine four different model hypotheses and use the same notation:
\begin{enumerate}
\item H1 : Data  is consistent with Gaussian errors, given by the measured uncertainties $\sigma_i$ ;
\item H2 : Same as H1, but data contains an additional unknown offset ($\sigma_{sys}$) ;
\item H3 : Data is described by Eq~\ref{eq};
\item H4 : Same as H3, with an additional systematic offset in the data.
\end{enumerate}

For each of the above hypothesis, we perform a maximum likelihood-based parameter estimation. Our maximum likelihood (valid when the dependent variables contain no errors) can be written as~\cite{astroml}:
\begin{equation}
{\cal L} = \prod_{i=1}^{n}\frac{1}{\sigma_i\sqrt{2\pi}}\exp\left\{-\frac{y_i-f(t,\mu,P,\phi_0)}{\sqrt{2} \sigma_i}\right\}^2,
\label{eq:lk}
\end{equation}
where $f(t,\mu,P,\phi_0)$ is given by Eq.~\ref{eq} for H3 and H4, else is equal to a constant offset.  For H1 and H3, $\sigma_i$ denotes the measured errors in $G$, whereas for H2 and H4, $\sigma_i$ is the quadrature sum of the measured uncertainties in $G$ and an unknown systematic term $\sigma_{sys}$.

We use the same data for our tests as P15 (a detailed documentation of his analysis can be found on {\tt github\footnote{https://github.com/mattpitkin/periodicG/}}). In all, this consists of 12 measurements of $G$ from 1981 to 2013. Similar to P15, we have excluded the measurements by  Karagioz \& Izmailov. References to all the other measurements can be found in Refs.~\cite{Anderson15a} and ~\cite{Mohr}.

\subsection{Model Comparison without errors in measurement times}
We now describe the results of our analysis without considering the errors in the measurement times. Results after including the errors shall be described in the next sub-section.
The first step in model comparison involves finding the best-fit parameters for each of the four hypotheses by maximizing
Eq.~\ref{eq:lk} for the pertinent model. Naively, one might select the best model as the one  with the largest value of the likelihood. But in model selection, one also needs to account for the different numbers of free parameters in each model. In Bayesian statistics, this is usually done by comparing the  model posteriors, as discussed  and implemented in P15. Other techniques  involve the use of penalized likelihoods such as  Akaike information criterion, Bayesian information criterion, etc~\cite{Shafer}. However, the results of any Bayesian model comparison test depends upon the choice of prior used for the model or the priors for parameters within each model. 

Therefore, to complement the Bayesian model comparison tests done in P15, we perform a frequentist model comparison. We assume that for the correct model, the data is normally distributed around the best-fit model with variance $\sigma_i^2$. Therefore, the sum of squares of  the normalized residuals around the best-fit model should follow $\chi^2$ distribution for the correct model after including the degrees of freedom for each model~\cite{astroml,Press92}. After finding  the best-fit model parameters for each hypothesis, we compare the  chi-square probability for the total degrees of freedom ($\nu$), given by~\cite{astroml,Press92},
\begin{equation}
P(\chi^2,\nu)=\frac{(\chi^2)^{(0.5\nu - 1)} \exp(-0.5\chi^2)}{2^{0.5\nu} \mathrm{\Gamma} (0.5\nu)},
\end{equation}
where for each model $f(x)$, 
\begin{equation}
\chi^2=\sum \limits_{i=1}^N \frac{\left(y_i -f(x)\right)^2}{\sigma_i^2}.
\end{equation}
The best-fit model is the one with the largest value of $P(\chi^2,\nu)$.
 We use the {\tt Amoeba}~\cite{Press92} minimization technique and as initial starting guesses for the model parameters in H3 and H4, we used the best-fit values found  by Anderson et al. Note that for H3 and H4, in all there are only 12 data points and four (five) free parameters and so we expect a number of false minima for the four (five) dimensional parameter space (See for eg.~\cite{Schlamminger}). We do not explore these false minima here, as our main goal is to test the claims in Ref.~\cite{Anderson15a}. 

\begin{figure}
\includegraphics[width=0.52\textwidth]{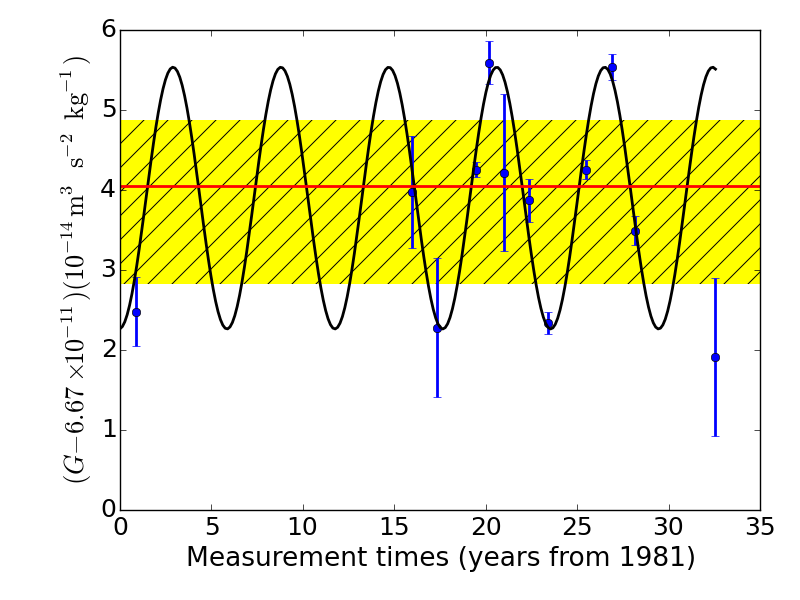}
\caption{Data showing variation of Newton's Constant ($G$) as a function of time. The blue points indicate the measurements (same as those used in P15~\cite{Pitkin15}, which contains links to original references). The red line shows best-fit constant offset for hypothesis H1. The yellow hatched region indicates best-fit for hypothesis H2. The black sinusoidal curve shows the best-fit parameters for hypothesis H3.}
\label{fig1}       
\end{figure}

Our results for all the four models can be found in Tab.~\ref{tab:bestfit}. The measurements for $G$ along with the best-fit model parameters for the hypotheses H1, H2, and H3 are shown in Fig.~\ref{fig1}.
We can see that the best-fit model is hypothesis H2, which consists of a constant offset and unknown systematic noise component. It is a better description of the data compared to sinusoidal variations. Our results were obtained by a complementary model comparison method. 
We conclude that the measurements of $G$ do not support a sinusoidal variation, without allowing for any errors in the measurement times, in agreement with P15.

\begin{table*}[t]
\caption{Summary of model comparison tests without including the errors in measurement for each of the four hypotheses H1, H2, H3, and H4. $A$, $\mu$, $P$, $\sigma_{sys}$ are defined in Eq.~\ref{eq}. For each of the hypotheses, 
we show  $\chi^2/DOF$ and P($\chi^2$,$\nu$). As we can see, P($\chi^2$,$\nu$) is largest for H2 (which is a constant offset and an unknown systematic). Therefore, the measurements do not favor a sinusoidal variation compared to a constant offset and a systematic term.}
\label{tab:bestfit}       
\begin{tabular}{lllllllll}
\hline\noalign{\smallskip}
Hypothesis & $\mu$ & $\sigma_{sys}$ & $A$ & $P(yrs)$ & $\phi_0$ & DOF & $\chi^2/DOF$ & P($\chi^2$,$\nu$)  \\
\noalign{\smallskip}\hline\noalign{\smallskip}
H1 & $6.766\times 10^{-11}$ & - & - & - & -  & 11 & 28.04 & $6.8 \times 10^{-60}$ \\
H2 & $6.674 \times 10^{-11}$ & $10^{-14}$ & - & - & - &  10 & 1.27 & 0.059 \\
H3 & $6.674 \times 10^{-11}$ & - & $1.64 \times 10^{-14}$ & $5.9$ & $-0.07$ & 8 & 2.93 & 0.0011 \\
H4 & $6.571 \times 10^{-11}$ & $10^{-12}$ & $1.9 \times 10^{-14}$  & $7.57$ & 0.0011 & 7 & 1.71 & 0.032 \\ 
\noalign{\smallskip}\hline
\end{tabular}
\end{table*}

\begin{table*}[t]
\caption{Summary of model comparison tests after including errors in measurement times. The results for 
hypothesis H1 and H2 will be same as in Tab.~\ref{tab:bestfit} and are not shown in this table.}
\label{tab:bestfit2}       
\begin{tabular}{lllllllll}
\hline\noalign{\smallskip}
Hypothesis & $\mu$ & $\sigma_{sys}$ & $A$ & $P(yrs)$ & $\phi_0$ & DOF & $\chi^2/DOF$ & P($\chi^2$,$\nu$)  \\
\noalign{\smallskip}\hline\noalign{\smallskip}
H3 & $6.674 \times 10^{-11}$ & - & $1.64 \times 10^{-14}$ & $5.9$ & $-0.07$ & 8 & 2.93 & 0.0011 \\
H4 & $6.571 \times 10^{-11}$ & $10^{-12}$ & $1.9 \times 10^{-14}$  & $7.57$ & 0.0011 & 7 & 1.71 & 0.033 \\ 
\noalign{\smallskip}\hline
\end{tabular}
\end{table*}

\subsection{Model Comparison after including errors in measurement times}
We now redo the analysis by including the uncertainties in measurement times. We use the same values for these as in 
P15, which is   $\sigma_t$=0.25 years for all the measurements, except for JILA-10 and LENS-14 for which $\sigma_t$= 1 week. To include the uncertainties in the dependent variables, we follow the formalism by Weiner et al~\cite{Weiner}, which has been used in a number of astrophysical analyses from galaxy clusters to pulsars~\cite{Hoekstra,Desai15}. Briefly, the likelihood is the same as in  Eq.~\ref{eq:lk}.  Only $\sigma_i$ gets modified and the new uncertainty $\sigma_{net}$ is given by 
\begin{equation}
\sigma_{net}=\sqrt{\sigma_i^2+\left(\frac{\partial y}{\partial t}\right)^2 \sigma_t^2}. 
\end{equation}
Intuitively, we would expect the $\chi^2$/DOF to be smaller compared to any fitting done without errors in the dependent variable. The best-fit values, $\chi^2/dof$, and P($\chi^2,\nu$) for H3 and H4 are shown in  Tab.~\ref{tab:bestfit2}. As we can see,  even after including the errors in the dependent variable, the chi-square probabilities are smaller for H3 and H4 compared to H2. Therefore, the data still supports a constant offset and a systematic uncertainty compared to a sinusoidal variation, even after the inclusion of errors in measurement times.

%
%

\section{Conclusions}
The current consensus among the Physics and Astrophysics community  is that measurements of Newton's Gravitational constant ($G$)  have no time dependence or any correlations with environmental parameters. However, this paradigm has recently been challenged by   Anderson and collaborators~\cite{Anderson15a}. They  have detected sinusoidal variations in the measurements of Newton's constant ($G$) (compiled over the last 35 years from 1981) with a period of 5.9 years~\cite{Anderson15a,Anderson15b}. Similar periodic variations have also been observed in measurements of the length of the day~\cite{Holme}. Therefore, Anderson et al have argued there is some systematic effect in measurements of $G$ that is connected with the same mechanism, which causes variations in the length of the day. However, these results have been disputed by Pitkin~\cite{Pitkin15}, who has shown using  Bayesian model comparison and a suitable choice of priors for the different model parameters, that  a model with constant offset and a unknown systematic uncertainty fits the data better than any sinusoidal variations, with the Bayes factor between the two hypotheses having a value equal to    $e^{30}$. Therefore, the analysis  by Pitkin contradicts the claims by Anderson et al.

In this letter, we have carried out a complementary analysis of the same 
dataset to resolve the above conflicting claims between the two groups of authors.  We have performed  frequentist model comparison tests of the same measurements, both with and without  the errors in measurement times. We examined four hypotheses similar to that in Ref.~\cite{Pitkin15}: a constant offset; a constant offset augmented by an unknown systematic uncertainty; sinusoidal variations; and sinusoidal variations with an unknown systematic uncertainty. For each of these models, we found the best-fit parameters and then calculated the chi-square probabilities for each of these and chose the best model as the one with the largest chi-square probability. This is the standard procedure followed in  frequentist model comparison, which is complementary to the Bayesian model comparison analysis done by Pitkin.
 
We find in agreement with Pitkin that the best model is the one with a constant offset in measurements of $G$ along with an unknown systematic offset. Therefore, there is no evidence for any sinusoidal variations in the measurements of $G$.

\begin{acknowledgements}
We are grateful to Jake VanDerPlas for his thorough detailed notes on model comparison~\footnote{http://jakevdp.github.io/blog/2015/08/07/frequentism-and-bayesianism-5-model-selection/} which helped us do this analysis  and also to Matt Pitkin for detailed documentation of his analysis on {\tt github}.
\end{acknowledgements}

\bibliographystyle{spphys}       
\bibliography{template}   

%
%

\end{document}